\documentclass[11pt,twoside]{article}
\usepackage{asp2014}

\aspSuppressVolSlug
\resetcounters

\begin{document}

\title{Evidence of Resonant Mode Coupling in the Pulsating DB White Dwarf Star KIC~08626021}
\author{W. Zong$^1$, S. Charpinet$^1$ and G. Vauclair$^1$}
\affil{$^1$ CNRS, Universit\'{e} de Toulouse, UPS-OMP, IRAP, Toulouse 31400, France; 
\email{weikai.zong@irap.omp.eu}, \email{stephane.charpinet@irap.omp.eu}, \email{gerard.vauclair@irap.omp.eu}}

\paperauthor{W. Zong}{Author1Email@email.edu}{}{IRAP}{CNRS, Universit\'{e} de Toulouse, UPS-OMP}{Toulouse}{}{31400}{France}
\paperauthor{S. Charpinet}{stephane.charpinet@irap.omp.eu}{}{IRAP}{CNRS, Universit\'{e} de Toulouse, UPS-OMP}{Toulouse}{}{31400}{France}
\paperauthor{G. Vauclair}{gerard.vauclair@irap.omp.eu}{}{IRAP}{CNRS, Universit\'{e} de Toulouse, UPS-OMP}{Toulouse}{}{31400}{France}

\begin{abstract}
The $Kepler$ spacecraft provides new opportunities to search for long term frequency and amplitude modulations of 
oscillation modes in pulsating stars. We analyzed nearly two years of uninterrupted data obtained with this instrument 
on the DBV star KIC 08626021 and found clear signatures of nonlinear resonant mode coupling affecting several triplets. 
The behavior and timescales of these amplitude and frequency modulations show strong similarities with theoretical 
expectations. This may pave the way to new asteroseismic diagnostics, providing in particular ways to measure for the 
first time linear growth rates of pulsation modes in white dwarf stars.
\end{abstract}

\section{Introduction}

The $Kepler$ spacecraft is a magnificent instrument to search for long term frequency and amplitude
modulations of oscillation modes in pulsating stars. Among the 6 pulsating white dwarfs present in the $Kepler$
field, KIC~08626021 is the unique DB pulsator. It has a rotation period P$_{rot}\sim1.7$ days, estimated from
the observed frequency spacings of 3 $g$-mode triplets \citep{os11}. It has been observed by $Kepler$ for 23 months in
short cadence (SC) mode without interruption. Thus, it is a suitable candidate to investigate the resonant mode
coupling mechanisms that could induce long term amplitude and frequency modulations of the oscillation modes.
Such resonant couplings are predicted to occur in triplets where the rotationally shifted
components have frequencies $\nu_1$ and $\nu_2$ such that $\nu_1\,+\nu_2\,\sim\,2\,\nu_0$, where $\nu_0$ is the frequency of the central
component. The theoretical exploration of those mechanisms was extensively developed long before the era of
space observations \citep{bu95,bu97} but was almost interrupted more than a decade ago because of the
lack of clear observational evidence of such phenomena, due to the difficulty of capturing amplitude or
frequency variations that occur on months to years timescales from ground based observatories. Resonant
coupling within triplets was proposed for the first time as the explanation for the frequency and amplitude
long term variations observed in the GW Vir pulsator PG 0122+200 \citep{va11}. We present the analysis of 
KIC~08626021 in which two triplets exhibit amplitude and frequency variations during the 23-month of
observation. Such time modulations pave the way to new asteroseismic diagnostics, providing in particular
ways to measure for the first time linear growth rates of pulsation modes in white dwarf stars.

\section{Frequency and amplitude modulations}

The white dwarf star KIC~08626021 has been continuously observed by $Kepler$ since quarter Q10.1 up to
Q17.2. The high precision photometric data cover $\sim$\,684.2 days (23 months), with a duty circle of $\sim$\,89 \%. We
used a dedicated software, FELIX, to extract frequencies (details of the program can be found
in \citealt{ch10}). 
In this study, we concentrate on rotationally split triplets and investigate the variation of amplitude and frequency 
between components of these triplets and their relationship. We point out that the frequencies near 3682\,$\mu$Hz reported 
by \citet{os11} were in fact resolved into several close peaks with the 23-month light curve. 
It is therefore probably not a real triplet contrary to the other structures found at
4310 and 5073\,$\mu$Hz (see below). 
In order to study the variability with time of these modes,
we constructed a filter window covering 200 days and slid the filter window along the whole light curve by time
steps of 20 days, thus constructing a time-frequency diagram. We also prewhithened the frequencies "chunk by
chunk", i.e., the 23-month light curve of KIC~08626021 was divided into 20 chunks, each
containing 6-month of data except the last 3 chunks being at the end of the observations. The results for the
two triplets are discussed below.
\articlefigure[width=1.0\textwidth]{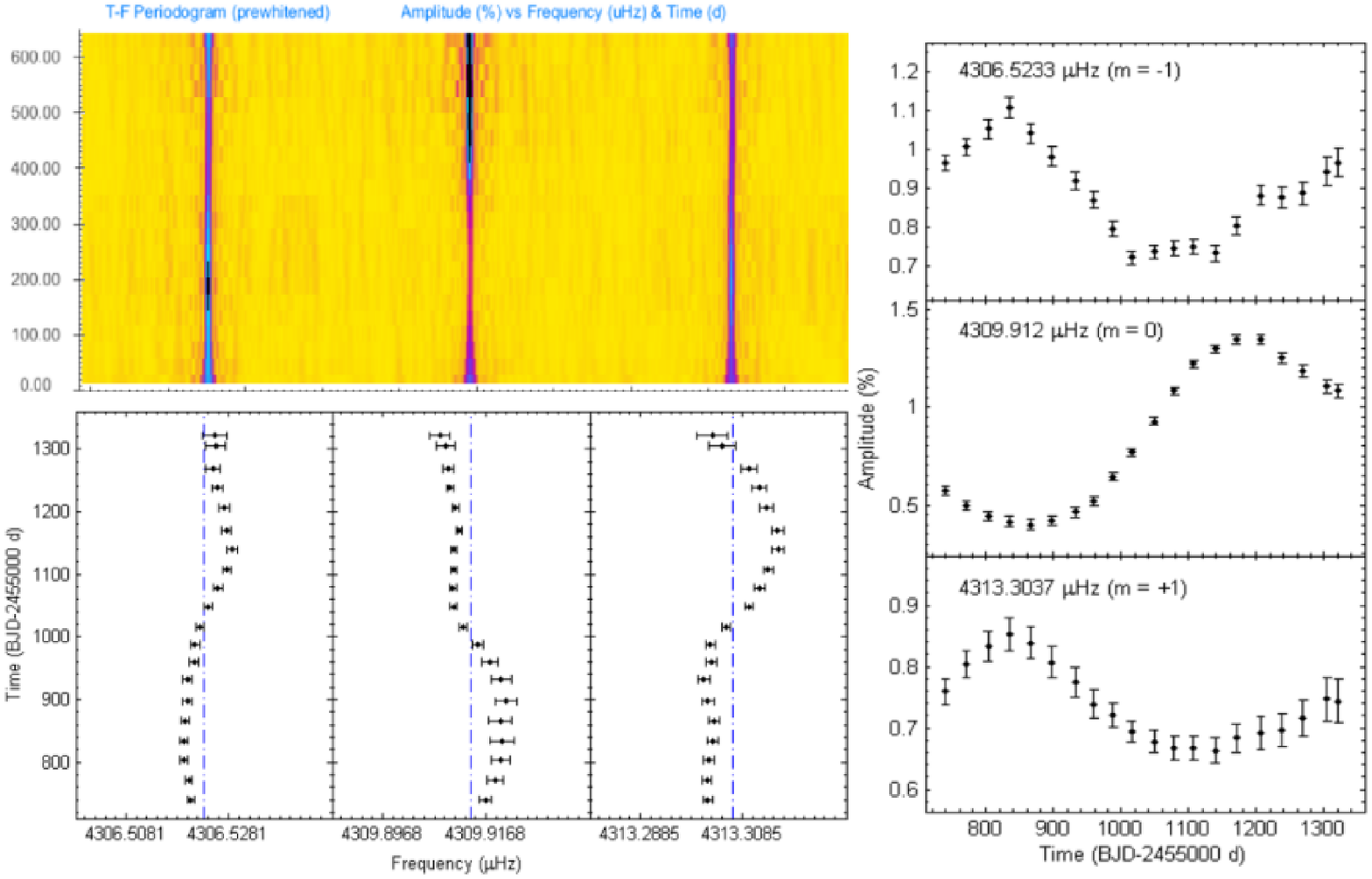}{pf_mod}{Frequency and amplitude modulations of the triplet at 4310\,$\mu$Hz. 
The frequencies and amplitudes
of the triplet show clear signatures of periodic modulations, as shown in the left bottom panel and right panel.
The grey scale in the upper left panel represents the amplitude. The dashed line in the lower right panel is
the average value of the frequency (see text for details). Note that the frequency scales
(x axis) in the upper and lower left panels are different.
}
\articlefigure[width=1.0\textwidth]{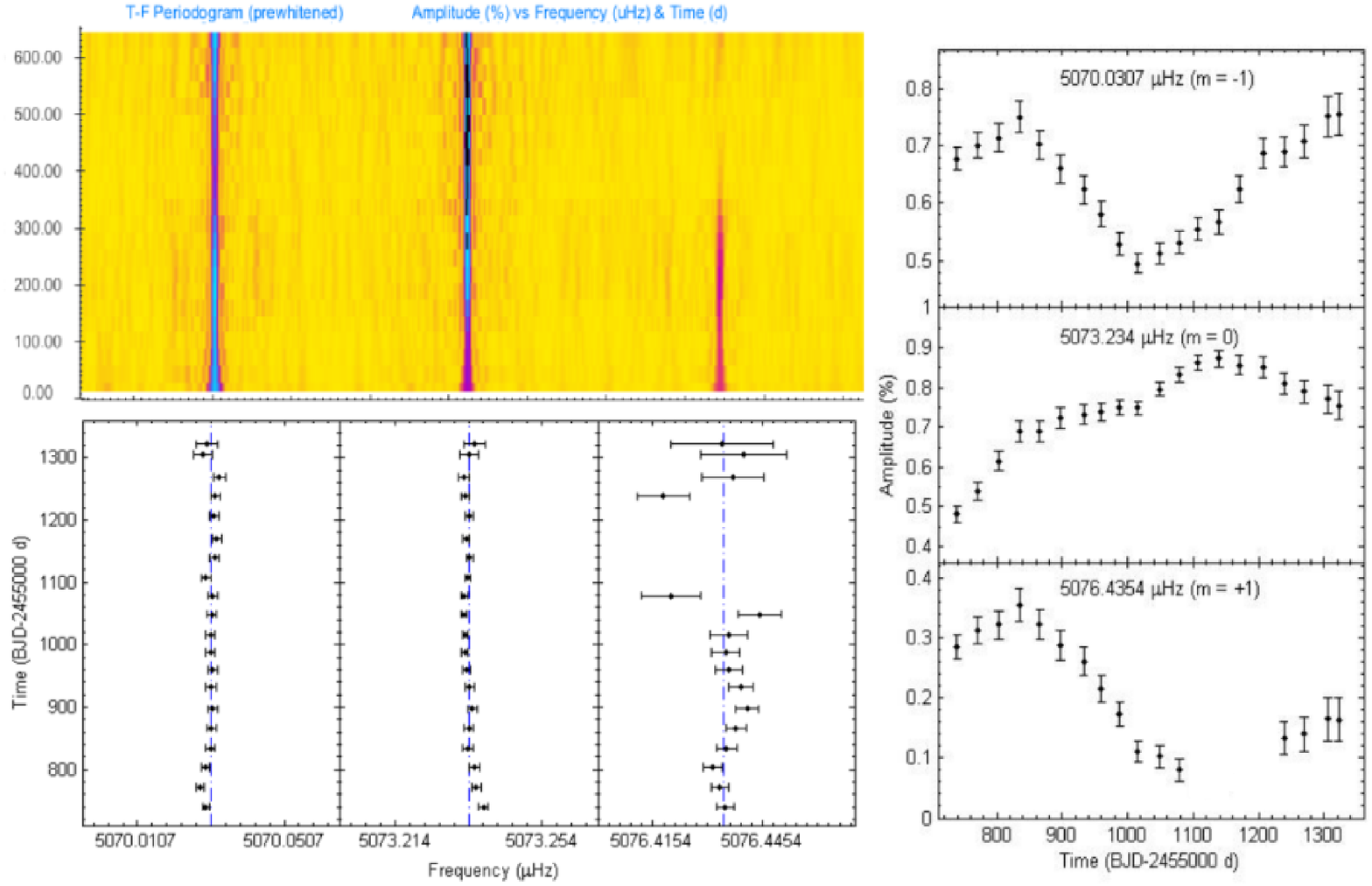}{sf_mod}{Frequency and amplitude modulations as shown in Figure\,\ref{pf_mod} 
but for the triplet at 5073\,$\mu$Hz.The amplitudes of the triplet show clear signatures of periodic modulations, as shown 
in the  right panel. The frequencies of the triplet are relatively stable during the observation.}

The amplitude and frequency modulations of the triplet near 4310 $\mu$Hz are shown in Figure\,\ref{pf_mod} . 
The grey scale (or color scale for electronic version) in the
upper left panel represents the amplitude. The vertical dashed line in the lower left panel is the average value of
the frequency over the entire run. The right panel shows the amplitude modulations of each component
forming the triplet. Both the amplitudes and frequencies show clear signatures of quasi periodic modulations with
the same timescale of $\sim$\,750 days. The frequencies and amplitudes of the side components evolve in phase and
are antiphased with the central component. Figure\,\ref{sf_mod} shows the modulations observed in the other triplet 
at 5073\,$\mu$Hz. The frequencies in this triplet appear to be stable during the nearly two years of monitoring,
while the amplitudes show modulations. 
Note that the amplitude of the $m$ = -1 component went down below the 4$\sigma$ detection threshold and was essentially 
lost in the noise during the last half of the observations.

\section{Discussion}
The frequency and amplitude modulations observed in the triplets of KIC 086226021 can be related to
nonlinear resonant mode coupling mechanisms. The first triplet at 4310\,$\mu$Hz (Figure\,\ref{pf_mod}) behaves like if it is in
the intermediate regime of the resonance, in which the oscillation modes undergo periodic amplitude and
frequency modulations. Theory suggests that the time scale of these modulations should be roughly a few
times the inverse of the growth rate of the pulsating mode \citep{go98}. Therefore this periodicity could in principle be
used to measure the growth rate.
In addition if we compare the second order effect of rotational splitting as
estimated directly from the measured mean frequencies :
\begin{displaymath}
\delta\nu = \nu_1 + \nu_2 - 2 \nu_0 \qquad ~\textrm{(1)}
\end{displaymath}
with the estimated value following \citet{dz92},
\begin{displaymath}
~~\delta\nu = 4 C \frac{\Omega^2}{\nu_0} ~~~~~~~~~~~~\qquad \textrm{(2)}
\end{displaymath}
where $C$ is the first order Ledoux constant ($\sim 0.5$ for dipole $g$-modes) and $\Omega\,=\,2\pi/P_{rot}$
is the angular frequency of the stellar
rotation, i.e ., $\delta\nu$\,$\sim$\,0.018\,$\mu$Hz, both are found to be very similar. In the intermediate regime, the expected
periodic modulation timescale is $P_{mod} \sim 1/\delta\nu$ \citep{go98}, which with the values given above leads to $P_{mod}\,\sim\,
650$ days, i.e., very similar to the amplitude and frequency modulation timescale of 750 days roughly estimated
from Figure\,\ref{pf_mod}. This further supports our interpretation that nonlinear resonant coupling is indeed at work in this
star. The triplet at 5073\,$\mu$Hz (Figure\,\ref{sf_mod}) would be in a different regime, likely in the nonresonant regime. The
amplitude of the components show clear modulations while the frequency are relatively stable during the
observation. This means the ratio of the real part over the imaginary part of the coupling coefficients is large in that case.
This ratio roughly measures nonlinear nonadiabaticities in the star. Hence our result shows that two neighbor
triplets can belong to different resonant regimes (frequency lock, time dependent or nonresonant), as it was
also suggested in the white dwarf star GD 358 \citep{go98}.

\section{Conclusion}
Frequency and amplitude modulations of oscillation modes have been found in several rotationally split
multiplets detected in the DB pulsator KIC 08626021, thanks to the high quality and long duration
photometric data obtained with the $Kepler$ spacecraft. These modulations show signatures pointing toward
nonlinear resonant coupling mechanisms occuring among the multiplet components. This is the first time that
such signatures are identified so clearly in white dwarf pulsating stars. Periodic modulations of frequency and
amplitude that occur in the intermediate resonant regime may allow for new asteroseismic diagnostics,
providing in particular a way to measure for the first time linear growth rates of pulsation modes in white
dwarf stars. Such results should motivate further theoretical work on nonlinear resonant mode coupling
mechanisms and revive interest in nonlinear stellar pulsation theory in general. Finally, we 
mention that similar modulations are also found in hot B subdwarf stars according to $Kepler$ data.






\acknowledgements WKZ acknowledges the financial support from the China Scholarship Council. 
This work was supported in part by the Programme National de Physique Stellaire (PNPS, CNRS/INSU, France) 
and the Centre National d'Etudes Spatiales (CNES, France).


\end{document}